\documentclass[reprint, superscriptaddress, secnumarabic, amssymb, nobibnotes, aps, prl]{revtex4-1}

\usepackage{graphicx}
\usepackage{epstopdf}
\usepackage[T1]{fontenc}
\usepackage[latin9]{inputenc}
\usepackage{amsbsy}
\usepackage{gensymb}
\usepackage[T1]{fontenc}
\usepackage[latin9]{inputenc}
\usepackage{amsmath}
\usepackage{amssymb}
\usepackage{bbm}
\usepackage{braket}
\usepackage{xcolor}
\allowdisplaybreaks
\usepackage{graphicx}
\usepackage{orcidlink}
  
\hypersetup
{
    bookmarks=true,         
    unicode=false,          
    pdftoolbar=true,        
    pdfmenubar=true,        
    pdffitwindow=false,     
    pdfstartview={FitH},    
    pdftitle={Broken time-reversal symmetry in a new  non-centrosymmetric superconductor Re$_{8}$NbTa},    
    pdfauthor={R. K. Kushwaha, R. P. Singh},     
    pdfsubject={},   
    pdfcreator={},   
    pdfproducer={}, 
    pdfkeywords={} {} {}, 
    pdfnewwindow=true,      
    colorlinks=true,       
    linkcolor=blue, 
    citecolor=blue,        
    filecolor=magenta,      
    urlcolor=blue           
}
\usepackage[normalem]{ulem}

\newcommand{\equref}[1]{Eq.~(\ref{#1})}

\newcommand{\figref}[1]{Fig.~\ref{#1}}

\newcommand{\tableref}[1]{Table~\ref{#1}}

\renewcommand{\approx}{\simeq}

\begin{document}

\title{\textrm{Broken time-reversal symmetry in a new non-centrosymmetric superconductor Re$_{8}$NbTa}}
\author{{R.~K.~Kushwaha}\,\orcidlink{0009-0005-3457-3653}}
\affiliation{Department of Physics, Indian Institute of Science Education and Research Bhopal, Bhopal, 462066, India}
\author{{Arushi}\,\orcidlink{0000-0003-4400-5260}}
\affiliation{Department of Physics, Indian Institute of Science Education and Research Bhopal, Bhopal, 462066, India}
\author{{S.~Sharma}\,\orcidlink{0000-0002-4710-9615}}
\affiliation{Department of Physics and Astronomy, McMaster University, Hamilton, Ontario L8S 4M1, Canada}
\author{S.~Srivastava}
\affiliation{Department of Physics, Indian Institute of Science Education and Research Bhopal, Bhopal, 462066, India}
\author{P.~K.~Meena}
\affiliation{Department of Physics, Indian Institute of Science Education and Research Bhopal, Bhopal, 462066, India}
\author{M.~Pula}
\affiliation{Department of Physics and Astronomy, McMaster University, Hamilton, Ontario L8S 4M1, Canada}
\author{J.~Beare}
\affiliation{Department of Physics and Astronomy, McMaster University, Hamilton, Ontario L8S 4M1, Canada}
\author{J. Gautreau}
\affiliation{Department of Physics and Astronomy, McMaster University, Hamilton, Ontario L8S 4M1, Canada}
\author{A.~D.~Hillier}
\affiliation{ISIS Facility, STFC Rutherford Appleton Laboratory, Didcot OX11 0QX, United Kingdom}	
\author{{G.~M.~Luke}\,\orcidlink{0000-0003-4762-1173}}
\affiliation{Department of Physics and Astronomy, McMaster University, Hamilton, Ontario L8S 4M1, Canada}
\affiliation{TRIUMF, Vancouver, British Columbia V6T 2A3, Canada}	
\author{{R.~P.~Singh}\,\orcidlink{0000-0003-2548-231X}}
\email[]{rpsingh@iiserb.ac.in} 
\affiliation{Department of Physics, Indian Institute of Science Education and Research Bhopal, Bhopal, 462066, India}

\begin{abstract}
\begin{flushleft}
\end{flushleft}
 Re-based superconductors provide a rich platform for the study of unconventional superconductivity. We have investigated the superconducting properties of Re$_{8}$NbTa, a new noncentrosymmetric cubic ($\alpha$-Mn structure) rhenium-based ternary superconductor using transport, magnetization, specific heat, and muon spin rotation/relaxation ($\mu$SR) measurements. Specific heat and transverse field $\mu$SR measurements suggest moderately coupled fully gapped superconductivity, well described by BCS theory. However, our zero-field $\mu$SR measurements reveal a small internal field onsetting around the superconducting T$_c$, indicating that the superconducting order parameter breaks the time-reversal symmetry.
\end{abstract}
\maketitle
\section{Introduction}
An important question in the study of unconventional superconductivity is the role of fundamental symmetries such as time reversal and inversion in the pairing interaction and the resulting superconducting state~\cite{Smidman, CePt3Si}. The discovery of the non-centrosymmetric (NCS) superconductor CePt$_{3}$Si~\cite{CePt3Si} attracted significant attention due to the absence of an inversion center in its crystal structure. This lack of inversion symmetry disrupts the degeneracy of electrons on the Fermi surface through antisymmetric spin-orbit coupling (ASOC)~\cite{ASOC}, resulting in the intriguing phenomenon of parity mixing~\cite{Smidman, TRSBreview}. These parity-mixed Cooper pairs have unveiled a plethora of captivating phenomena in NCS superconductors, such as a high upper critical field surpassing the Pauli limiting field~\cite{Pauli1,Pauli2}, the emergence of a helical superconducting phase in the presence of a magnetic field (referred to as the FFLO state)~\cite{FF,LO}, the presence of nodes or multigap behavior~\cite{nl1,nl2,nl3,nl4}, the existence of topological superconducting phases~\cite{TSC}, and the recent revelation of superconducting torofluxes, reminiscent of the Chandrasekhar-Kendall state~\cite{CK}. In addition to these unconventional properties, some NCS superconductors have been reported to show spontaneous breaking of the time-reversal symmetry (TRS) within the superconducting state~\cite{TRSBreview}.\\ 
In recent years, extensive research has been focused on noncentrosymmetric (NCS) superconductors~\cite{NbOs, TaOs, NbOs2, AuBe, Zr3Ir, Zr2Ir, Re3Ta, Re3W1, Re3W2, Re6Hf, Re6Ti, Re6Zr, ReB, ReTRSB, Re5.5Ta, Re0.82Nb0.18, L7I3, L7R3, L7N3, NbReSi}, with a particular focus on rhenium-based NCS superconductors due to the frequent occurrence of time-reversal symmetry (TRS) breaking~\cite{Re6Hf,Re6Ti,Re6Zr,ReTRSB}. However, despite crystallization in the NCS structure, some of these rehinium-based superconductors exhibit preserved TRS~\cite{Re3Ta,Re3W1,Re3W2,ReB,Re5.5Ta}, while centrosymmetric rhenium-based superconductors and elemental rhenium break TRS~\cite{ReTRSB} questioning the role of crystal structure in TRS breaking. Furthermore, a study focused on the Re$_{1-x}$Mo$_{x}$ series~\cite{ReMo} compounds crystallized in centro/noncentrosymmetric crystal structure introduced the intriguing hypothesis that the ratio of rhenium to the transition metal (X) and the optimal percentage of rhenium could be factors that contribute to TRS breaking~\cite{ReMoTRSB}. These observations have prompted inquiries into the potential roles of ASOC, its strength~\cite{Re6Zr,TaXSi,LaNiC2}, and crystal structure, in the context of TRS breaking in rhenium-based superconductors. Further research and microscopic measurements on new rhenium-based superconductors are necessary to fully understand the exact pairing mechanism. The ternary Re-based alloy, Re$_{8}$NbTa, offers an excellent opportunity for this purpose due to its mixed 4d/5d site. This study also probes the role of disorder in TRSB in rhenium-based systems, an aspect which remains largely unexplored. Compared to the binary analog, Re$_{8}$NbTa introduces additional disorder due to the slight variation in the atomic size of Ta and Nb. This difference is evident in the lower residual resistivity ratio of Re$_{8}$NbTa compared to binary Re-based alloys~\cite{Re5.5Ta,Re0.82Nb0.18}.\\
In this paper, we provide the first microscopic investigation of the rhenium-rich ternary superconductor Re$_{8}$NbTa, characterized using AC transport, magnetization, specific heat, and zero/transverse field muon spin relaxation/rotation ($\mu$SR)~\cite{Muon} measurements. Our transverse field $\mu$SR measurements indicate a fully-gapped isotropic superconducting state, consistent with an s-wave pairing state~\cite{BCS}. Intriguingly, our zero field $\mu$SR results reveal time-reversal symmetry breaking (TRSB) in the superconducting state, indicating an unconventional nature to superconductivity in Re$_{8}$NbTa.

\section{Experimental Details}
\begin{figure}  [t!] 
\includegraphics[width=0.97\columnwidth, origin=b]{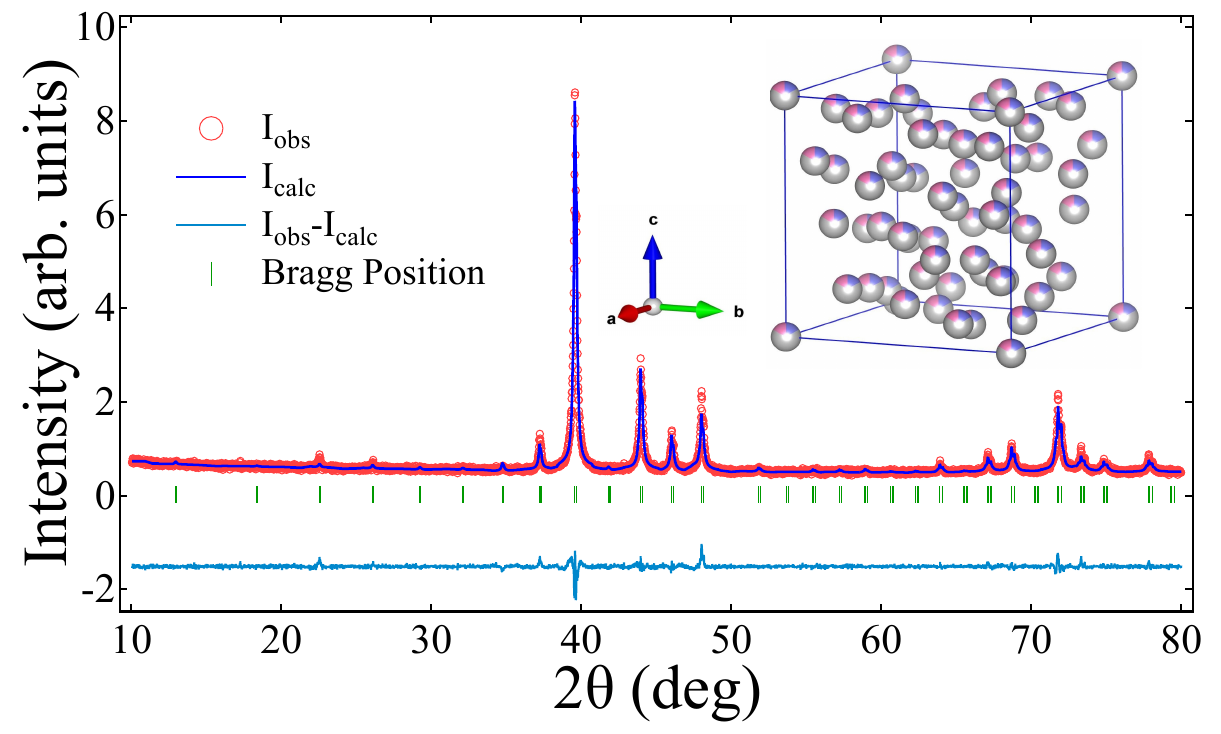}
\caption{\label{XRD} The powder XRD pattern of Re$_{8}$NbTa is shown by pink circles, while the solid blue line represents the Rietveld refinement pattern. The inset shows the crystal structure of Re$_{8}$NbTa}
\end{figure}
The polycrystalline Re$_{8}$NbTa sample was synthesized by melting high purity metals (3N) Re, Nb and Ta in a stoichiometric ratio under high-purity argon gas in an arc furnace. A PANalytical powder X-ray diffractometer equipped with Cu-K$_{\alpha}$ radiation ($\lambda$ = 1.5406~\AA) was used to determine the crystal structure and phase purity. Magnetization was measured with a Quantum Design MPMS 3, while resistivity and specific heat were measured using a Quantum Design PPMS. Muon spin rotation/relaxation experiments were conducted in transverse-field (TF) and zero-field (ZF) configurations to investigate the superconducting gap structure and identify any spontaneous field in the superconducting state, using the M20 beamline at the TRIUMF Center for Molecular and Material Science (CMMS), in Vancouver, Canada. For the $\mu$SR measurements, the Re$_{8}$NbTa sample was cut into flat plates and held in the sample holder within the $^4$He gas flow cryostat using thin metalized Mylar$^{\tiny{\textcircled{R}}}$ tape, which is sufficiently thin to not stop any muons. Muons missing the sample landed in a veto detector, allowing us to collect background-free spectra. In the TF configuration, the initial orientation of the muon spins was prepared to be perpendicular to the magnetic field, which was applied parallel to the incoming muon beam momentum. A combination of a fluxgate magnetometer, muMetal, and three sets of perpendicular electromagnets was used to establish a zero magnetic field at the sample position. Our $\mu$SR data was analyzed using the musrfit software package~\cite{MuSRfit}.

\section{Results and Discussion}
\begin{figure} [t!] 
\includegraphics[width=0.97\columnwidth, origin=b]{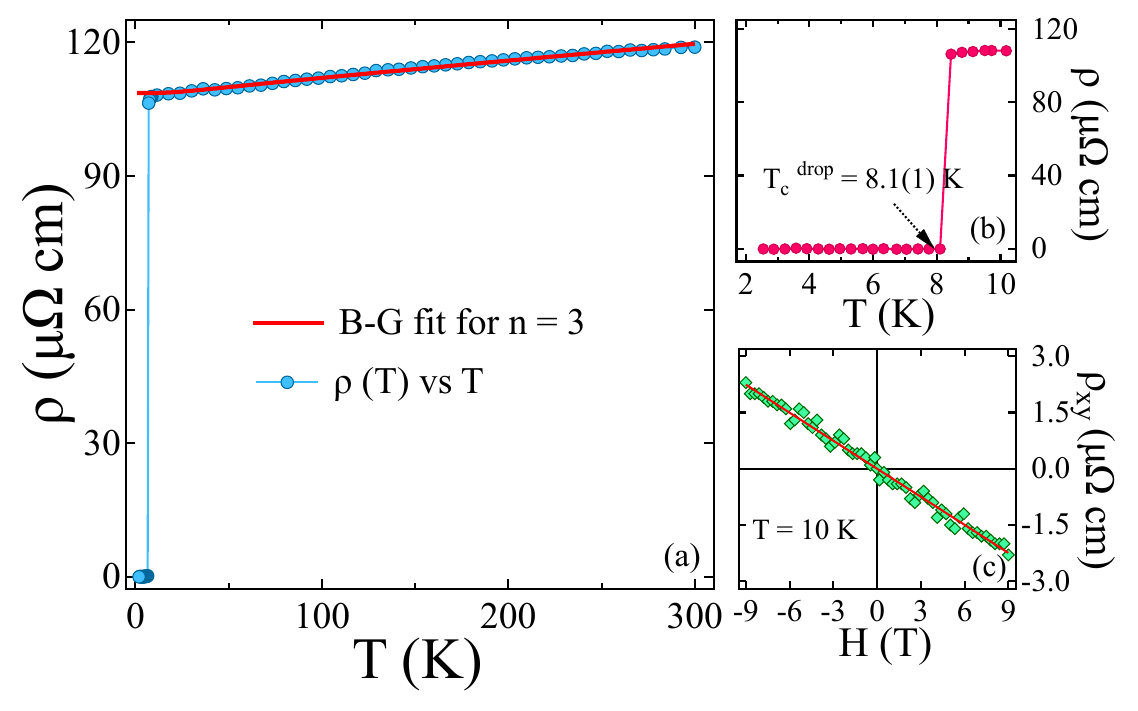}
\caption{\label{rho}(a) Temperature dependence of zero -field resistivity, $\rho(T)$. The red line represents the BG fit (b) A sharp zero drop in resistivity at $T_{c} ^{drop}$ = 8.1(1)~K. (c) Hall resistivity ($\rho_{xy}$) at 10~K under a magnetic field of $\pm$9~T.}
\end{figure}

\subsection{Crystal Structure Characterization}
The room-temperature powder X-ray diffraction (XRD) pattern of Re$_{8}$NbTa is shown in \figref{XRD}. Rietveld refinement using Fullprof software~\cite{Fullprof} confirms the phase purity and crystal structure as $\alpha$-Mn cubic with space group I$\Bar{4}3m$. The lattice parameters and cell volume obtained from the refinement are as follows: a = b = c = 9.654(9)~\text{\AA} and $V_{cell}$ = 899.8(9)~\text{\AA}$^{3}$.
\begin{figure*}[t!]
\includegraphics[width=1.99\columnwidth,origin=b]{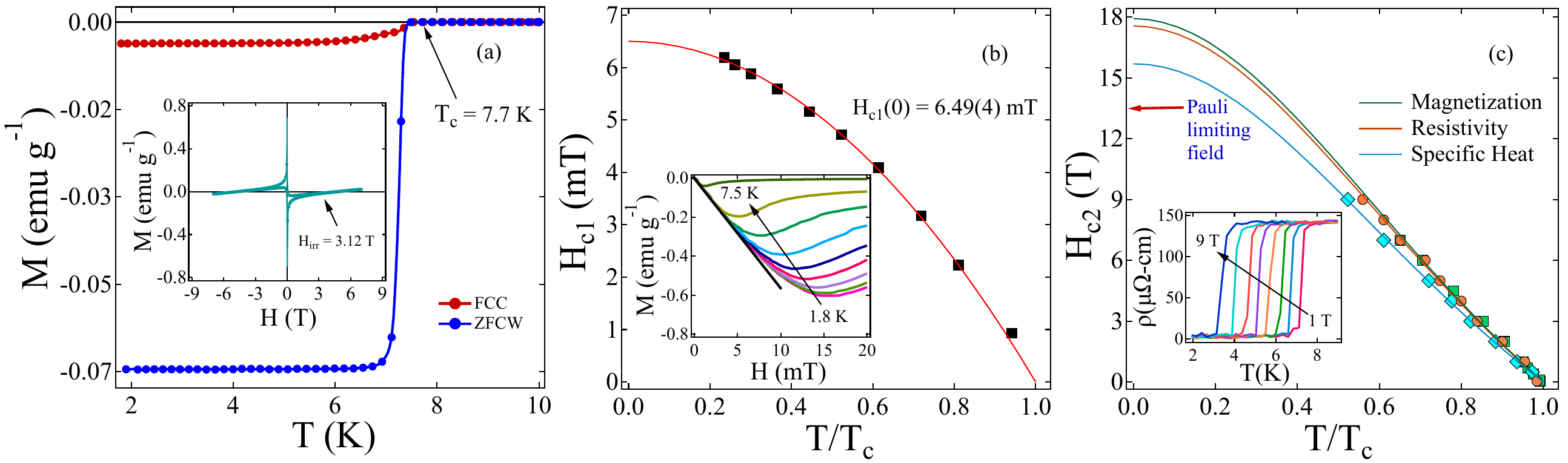}
\caption{\label{Mag}(a) The temperature dependence of the moment in the ZFCW and FCC mode under an applied magnetic field of 1 mT. The inset shows the M(H) loop at 1.8~K under an applied field of $\pm$7~T. (b) Lower critical field vs reduced temperature ($T / T_{c}$). The inset provides M(H) curves at various temperatures. (c) The upper critical field vs the reduced temperature ($T/T_{c}$), estimated from magnetization, resistivity, and specific heat measurements. Solid lines represent the Ginzburg-Landau (GL) fit using equation \equref{Hc2}, with the inset showing the variation of resistivity with temperature in various magnetic fields}.
\end{figure*}
\subsection{Electrical Resistivity}
Figure \ref{rho}(a) shows the temperature-dependent resistivity ($\rho(T)$) of Re$_8$NbTa, measured from 1.9 to 300~K in zero magnetic field. Figure \ref{rho}(b) shows a distinct decrease in resistivity at critical temperature $T_c^{drop} = 8.1(1)$~K, with a transition width of 0.5~K, which, along with magnetization and specific heat measurements discussed later, confirms bulk superconductivity in Re$_8$NbTa. The residual resistivity ratio (RRR = $\rho(300 K)$ /$\rho(10 K)$) = 1.1 implies a highly disordered nature and is in agreement with other Re-based alloys crystallizing in the $\alpha$-Mn structure~\cite{Re6Zr,Re6Hf,Re6Ti,Re5.5Ta}. $\rho(T)$ exhibits almost temperature-independent behaviour in the normal state region, which allows a fit to these data employing the Bloch-Gr\"{u}neisen (BG) model~\cite{BG}. In such a case, the temperature dependence of the resistivity is analysed using \equref{res} (in our case, n = 3)
\begin{equation}
\rho(T)=\rho_{0}+B \left({\frac{T}{\Theta_{D}}}\right)^{n}\int_{0}^{\Theta_{D}/T}\frac {x^{n}}{(e^{x}-1)(1-e^{-x})}dx,
\label{res}
\end{equation}
where $\rho_{0}$ represents the temperature-independent residual resistivity due to scattering from defects in the crystal; the second term is the temperature-dependent generalized BG resistivity in which $B$ is a material-dependent proportionality constant. The red line in \figref{rho}(a) shows the best fit to the data and yields $\rho_{0}$ = 108.6(3)~$\mu\Omega\cdot cm$, $B = 1.11(1)\;\mu\Omega\cdot cm$, and $\Theta_{D}$ = 149(1)~K.

\subsection{Magnetization}

\begin{figure*}[t!] 
\includegraphics[width=2.0\columnwidth,origin=b]{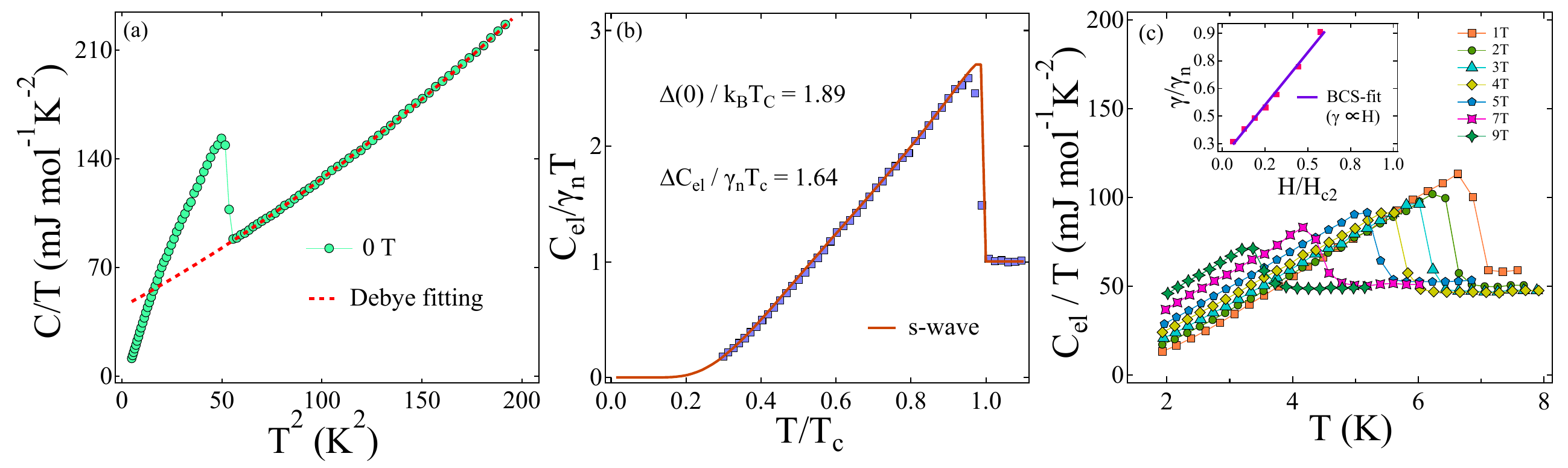}
\caption{\label{SH}(a) Zero-field $C/T$ variation to $T^{2}$ with the red dotted line representing the fit to \equref{C/T}. (b) Normalized electronic-specific heat is best described by the isotropic single-gap model (solid orange line). (c) $C_{el}/T$ vs $T$ at various magnetic fields; the inset indicates a BCS-type single gap structure as the field-dependent Sommerfeld coefficient is shown as a straight line.}
\end{figure*}

Magnetization measurements were performed in zero field-cooled warming (ZFCW) and field-cooled cooling (FCC) in an applied magnetic field of 1~mT. It confirms the superconducting transition temperature of Re$_{8}$NbTa at 7.7(4)~K by exhibiting a strong diamagnetic signal as shown in \figref{Mag}(a) and the inset shows the full range M-H curve from -7~T to 7~T. This magnetization loop defines the irreversibility field $H_{irr}$ = 3.1(2)~T, above which the unpinning of the vortices takes place. The lower critical field, $H_{c1}$(0), was determined from the low-field magnetization curves taken at different temperatures from 1.8 to 7.5 K, which are shown in the inset of \figref{Mag}(b). The lower critical field corresponding to every temperature curve was considered as the point that deviates from the linearity relation (shown by the solid black line in the inset of \figref{Mag}(b)). $H_{c1}$(0) was found to be 6.49(4)~mT from the fit utilizing the Ginzburg-Landau (GL) relation
\begin{equation}
H_{c1}(T)=H_{c1}(0)\left[1-\left(\frac{T}{T_{c}}\right)^{2}\right].
\label{Hc1}
\end{equation}
The upper critical field $H_{c2}$(0) was estimated by measuring the effect of the applied field on the transition temperature using different techniques: magnetization (M (T, H)), specific heat (C (T, H)) and resistivity ($\rho$ (T, H)). Figure \ref{SH}(c) shows (C (T, H)) where $T_{c}$ is suppressed only down to 3.54~K in the presence of a field of 9~T, suggesting a high value of the upper critical field. In M (T, H), the onset of the diamagnetic signal is considered as $T_{c}$, while the midpoint of the jump and the drop in C (T, H) and $\rho$ (T, H) (see the inset of \figref{Mag} (c)), respectively, are taken as criteria for $T_{c}$. $H_{c2}$(T) of all measurements was plotted against reduced temperature ($T/T_{c}$), and a linear response was observed near $T/T_{c}$ = 1, which can be well described by the GL relation for the upper critical field which follows
\begin{equation}
\label{Hc2}
H_{c2}(T) = H_{c2}(0)\left[\frac{1-t^{2}}{1+t^{2}}\right],
\end{equation}
where, $t = T/T_{c}$ and the fit using \equref{Hc2} yields $H_{c2}(0)$ = 17.9(2)~T, 17.5(3)~T and 15.7(4)~T from magnetization, resistivity and specific heat measurements, respectively. Such high values of $H_{c2}(0)$ were also found for other NCS Rhenium-based alloys~\cite{Re5.5Ta,Re6Zr}, which indicates the possibility of a strong ASOC and unconventional characteristic in the superconducting state of Re$_{8}$NbTa~\cite{CePt3Si,Smidman}.

Two independent Cooper pair-breaking processes can be considered with an external magnetic field application: the orbital pair-breaking effect and the Pauli limiting field effect. Orbital pair breaking occurs when the kinetic energy of one electron is increased so that the Lorentz force breaks the Cooper pair and the field at which the Cooper pair is broken, termed the orbital limiting field, which the Werthamer-Helfand-Hohenberg (WHH) model provides~\cite{WHH1,WHH2}
\begin{equation}
    H_{c2}^{orb}(0)= -\alpha T_{c} \left.\frac{dH_{c2}(T)}{dT}\right\vert_{T=T_{c}},
\end{equation}
At $T = T_{c}$, the initial slope $\frac{-dH_{c2}(T)}{dT}$ for Re$_{8}$NbTa was found to be 3.16(7)~T/K and by taking $\alpha$ = 0.693 (for dirty limit superconductors), orbital limiting upper critical field, $H_{c2}^{orb}(0)$ = 15.5(1)~T. In the realm of BCS superconductors, the Pauli limiting field is given by $H_{c2}^{P} = C \times T_{c}$, where $C$ = 1.86~T/K~\cite{Pauli1,Pauli2}, and for Re$_{8}$NbTa with $T_{c}$ = 7.7(4)~K, $H_{c2}^{P}$ = 14.3(2)~T. Both pair-breaking limiting upper critical fields are less than the estimated $H_{c2}(0)$ values obtained from three different measurements mentioned above, suggesting a possible unconventional superconducting nature in Re$_{8}$NbTa.

Two fundamental length scales of a superconductor, coherence length and penetration depth, can be computed using the values of $H_{c1}(0)$ and $H_{c2}(0)$. In GL theory, the coherence length $\xi_{GL}(0)$~\cite{Tinkham} and the penetration depth $\lambda_{GL}(0)$~\cite{lambda} were defined according to the following relations $H_{c2}(0) = \frac{\Phi_{0}}{2\pi\xi_{GL}^{2}}$ and $H_{c1}(0) = \frac{\Phi_{0}}{4\pi\lambda_{GL}^2(0)}\left( \ln \frac{\lambda_{GL}(0)}{\xi_{GL}(0)} + 0.12\right)$, respectively,
where $\Phi_{0}$ (= 2.07 $\times$ 10$^{-15}$~T m$^{2}$) is the magnetic flux quantum~\cite{Tinkham}. Using $H_{c1}(0)$ = 6.49(4)~mT and $H_{c2}(0)$ = 17.9(1)~T, $\xi_{GL}(0)$ and $\lambda_{GL}(0)$ were evaluated to be 42.8(7)~$\text{\AA}$ and 3371(2)~$\text{\AA}$, respectively. The GL parameter defined as $k_{GL}$ = $\frac{\lambda_{GL}(0)}{\xi_{GL}(0)}$ = 78.6(2), indicating Re$_{8}$NbTa is a strong type II superconductor. The thermodynamic critical field is estimated as $H_{c}(0)$ = 163.3(6)~mT using the relation $H_{c1}(0)H_{c2}(0)$ = $H_{c}^2\ln(k_{GL})$.

The Ginzburg-Levanyuk number $Gi$ represents the strength of the thermal fluctuations over the vortex unpinning in the superconducting state, and is given by \equref{Gi}
\begin{equation}
\label{Gi}
Gi= \frac{1}{2} \left(\frac{k_{B}\mu_{0} \tau T_{c}}{4\pi \xi(0)^{3} H_{c}^{2}(0)}\right)^{2},
\end{equation}
where the anisotropy factor $\tau$ is 1 for cubic Re$_{8}$NbTa. Using $T_{c}$ = 7.7(4)~K, $\xi(0)$ = 42.8(7)~\AA~ and $H_{c}(0)$ = 163.3(6)~mT, we obtained $Gi$ = 12.2(1) $\times$ 10$^{-6}$ which is greater than the typical values of low T$_{c}$ superconductors (~10$^{-8}$)~\cite{lowTc} but lesser than the value of high T$_{c}$ cuprate superconductors (~10$^{-2}$)~\cite{highTc}, suggesting that weak thermal fluctuations are responsible for the unpinning of vortices~\cite{GI} in Re$_{8}$NbTa.

\subsection{Specific Heat}\label{SHsection}

Figure \ref{SH}(b) shows the variation of normalized electronic specific heat with reduced temperature ($T/T_{c}$) in zero-field. A sharp jump at the $T_{c, mid}$ = 7.5(1)~K is aligning with the drop in resistivity and the onset of diamagnetism, confirms the bulk superconductivity in Re$_{8}$NbTa. In general, the normal state low-temperature specific heat data can be described by the Debye relation
\begin{equation}
C/T=\gamma_{n}+\beta_{3}T^{2}+\beta_{5}T^{4},
\label{C/T}
\end{equation}
where $\gamma_{n}$ represents the Sommerfeld coefficient providing electronic contributions to the specific heat, $\beta_{3}$ and $\beta_{5}$ represents the phononic and anharmonic contributions to the specific heat, respectively. A fit to the specific heat data using \equref{C/T} as shown in \figref{SH}(a), yields $\gamma_{n}$ = 44.5(4)~mJ mol$^{-1}$K$^{-2}$, which is relatively high as found in other intermetallic compounds like La$_{7}$X$_{3}$ (X= Ni, Ir) (44.43 ~mJ mol$^{-1}$K$^{-2}$ in La$_{7}$Ni$_{3}$~\cite{L7N3}, and  47~mJ mol$^{-1}$K$^{-2}$ in La$_{7}$Ir$_{3}$~\cite{L7I3}) due to moderately strong electronic correlation, $\beta_{3}$ = 0.695(6)~mJ mol$^{-1}$ K$^{-4}$ and $\beta_{5}$ = 1.30(2)~$\mu$J~mol$^{-1}$K$^{-6}$. $\gamma_{n}$ is related to the density of states at Fermi level $D_{c}(E_{F})$ by the relation $\gamma_{n}$ = $\left(\frac{\pi^{2}k_{B}^{2}}{3}\right)D_{C}(E_{f})$, where $k_{B}$ $\approx$ 1.38 $\times$ 10$^{-23}$~J K$^{-1}$. $D_{c}(E_{F})$ is estimated to be 18.9(1)~states/eV f.u. for Re$_{8}$NbTa. According to the Debye model, the Debye temperature ($\theta_{D}$) can be obtained from $\beta_{3}$ using the formula $\theta_{D}$ = $\left(\frac{12\pi^{4}RN}{5\beta_{3}}\right)^{1/3}$ where $N$ is the number of atoms per formula unit, $R$ is the molar gas constant (8.314~J~mol$^{-1}$K$^{-1}$), giving $\theta_{D}$ = 303(1)~K. It exceeds $\theta_{D}$ obtained from the BG fit of resistivity data, possibly due to the sensitivity of the low-temperature specific heat to low energy phonons, while the higher temperature resistivity responds to both low- and high-energy phonons~\cite{thetaDdiff}.

From McMillan's model~\cite{McM}, we can estimate the electron-phonon coupling strength from a dimensionless quantity $\lambda_{e-ph}$, which depends on the estimated value of $\theta_{D}$ and $T_{c}$ as
\begin{equation}
\lambda_{e-ph} = \frac{1.04 + \mu^{*}\ln(\theta_{D}/1.45T_{c})}{(1 - 0.62\mu^{*})\ln(\theta_{D}/1.45T_{c}) - 1.04}
\end{equation}
where $\mu^{*}$ represents screened Coulomb repulsion, considering $\mu^{*}$ = 0.13 as mentioned for all transition metals~\cite{McM}, $\theta_{D}$ = 303(1)~K and $T_{c}$ = 7.5(1)~K, we have obtained $\lambda_{e-ph}$ = 0.73, classifying Re$_{8}$NbTa as a strongly coupled superconductor.
The electronic contribution to the specific heat can be calculated after subtracting the phononic contribution ($C_{ph}$) from the total specific heat as $C_{el} = C - C_{ph} = C - \beta_{3}T^{3} - \beta_{5}T^{5}$. The normalized electronic specific heat jump $\Delta C_{el}/\gamma_{n}T_{c}$ = 1.64, higher than the standard value of 1.43 for BCS superconductors in the weak coupling limit, suggesting moderate electron-phonon coupling in Re$_{8}$NbTa~\cite{sep1,sep2}.

We have also performed field-dependent specific heat measurements to investigate the gap structure via the dependence of the Sommerfeld coefficient on the magnetic field ($\gamma(H)$). In a conventional superconductor $\gamma(H)$ $\propto$ $H$ for a nodeless gap~\cite{gamma2}. The Sommerfeld coefficient was calculated by fitting $C_{el}/T$ vs $T$ curves (as shown in \figref{SH}(c)) at different magnetic fields using the following relation~\cite{gamma}.
\begin{equation}
\frac{C_{el}}{T} = \gamma+\left(\frac{a}{T}\right) \exp{\left(-b\frac{T_{c}}{T}\right)}.
\label{gamma}
\end{equation}
The inset of \figref{SH}(c) represents the linear dependence of $\gamma/\gamma_{n}$ relative to the reduced magnetic field ($H/H_{c2}$) confirming the fully gapped superconductivity~\cite{gamma1,gamma2}.

The absence of noticeable temperature dependence in the electronic specific heat at low temperatures typically indicates a single, nodeless gap. Analyzing the temperature-dependent electronic specific heat allows us to explore the symmetry of the superconducting gap. Normalized entropy (S) in the superconducting region and electronic-specific heat can be related as $C_{el} = t~\frac{dS}{dt}$ where $t = T/T_{c}$ is the reduced temperature. Within the BCS approximation, the normalized entropy for a single gap is defined by the following relation
\begin{equation}
\frac{S}{\gamma_{n}T_{C}} = -\frac{6}{\pi^2}\left(\frac{\Delta(0)}{k_{B}T_{C}}\right)\int_{0}^{\infty}[ \textit{f}\ln(f)+(1-f)\ln(1-f)]dy,
\label{BCS}
\end{equation}
where, $\textit{f}(\xi) = \left[\exp(\textit{E}(\xi)/k_{B}T)+1\right]^{-1}$ is the Fermi function, $\textit{E}(\xi) = \sqrt{\xi^{2}+\Delta^{2}(t)}$, $\textit{E}(\xi)$ is the energy of the normal electrons measured relative to the Fermi energy, $\textit{y}$ = $\xi/\Delta(0)$, and $\Delta(t) = \tanh\left[1.82(1.018((\mathit{1/t})-1))^{0.51}\right]$ is the BCS approximation for temperature-dependent superconducting energy gap. The solid orange line in \figref{SH}(b) represents the fit to the electronic specific heat data with the single isotropic BCS gap model. We obtain a normalized gap value, $\Delta(0)/k_{B}~T_{c}$ = 1.89, slightly higher than the weakly coupled BCS value (1.76). This increase in both $\Delta C_{el}/\gamma_{n}T_{c}$ and $\Delta(0)/k_{B}T_{c}$ values compared to BCS values has also been observed for other Re-based binary alloys crystallizing in the $\alpha$-Mn structure~\cite{Re6Hf,Re6Ti,ReMo,Re5.5Ta}, indicating a moderate electron-phonon coupling in those materials and here for Re$_{8}$NbTa.
\begin{figure} [t!]
\includegraphics[width=0.98\columnwidth,origin=b]{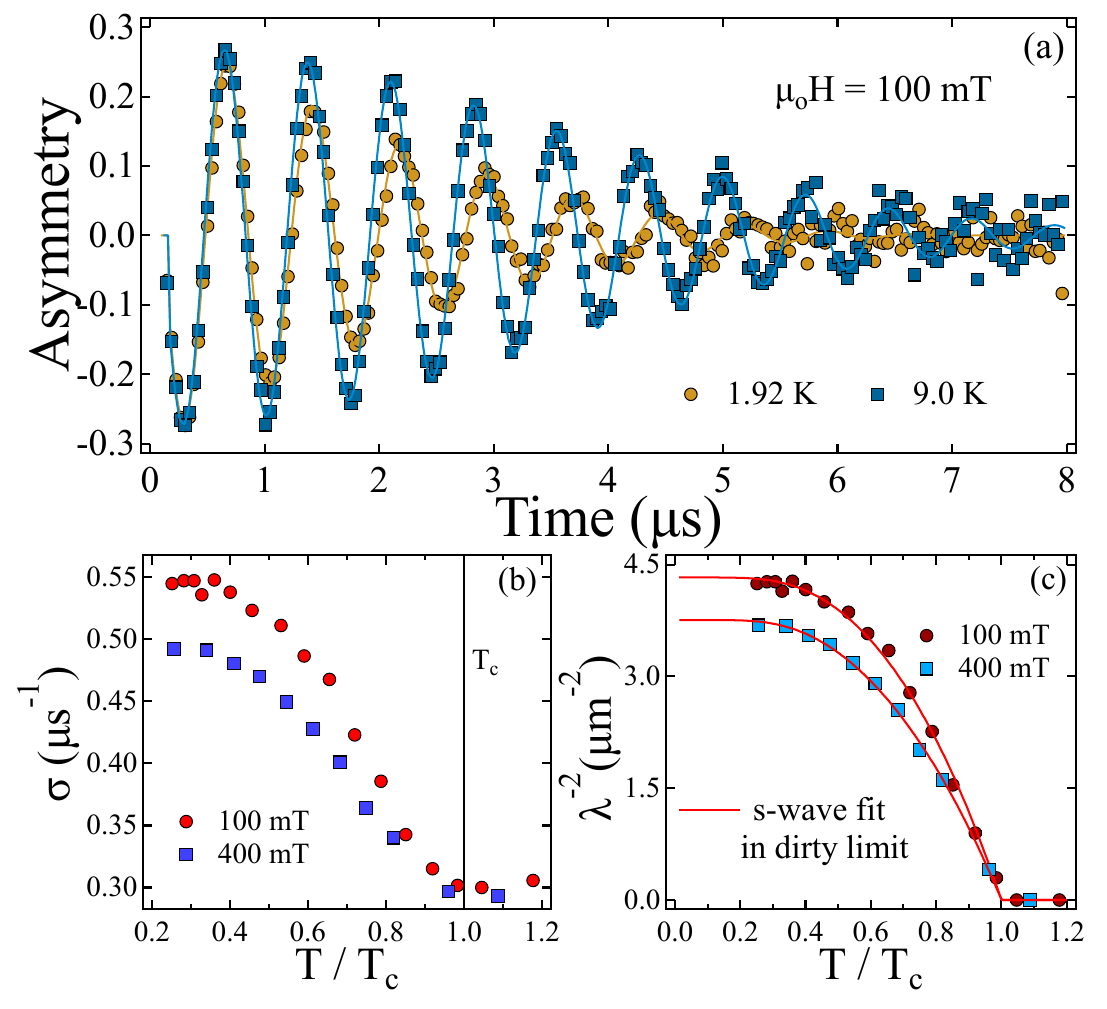}
\caption{\label{TF} (a) TF-$\mu$SR asymmetry spectra recorded from 1.8 to 10.0~K at 100~mT. Plots at 1.92~K and 9~K in a rotating reference frame of 90~mT highlight changes upon flux-line lattice formation in the superconducting state. Solid lines depict fits using \equref{ATF}. (b)  TF-$\mu$SR depolarization rate $\sigma$ at 100~mT and 400~mT vs. reduced temperature $T/T_c(H)$ (c) Inverse squared penetration depth vs. reduced temperature $T/T_c(H)$ in 100~mT and 400~mT for Re$_{8}$NbTa.}
\end{figure}

\subsection{Muon Spin Rotation/Relaxation}

\textit{Transverse-field $\mu$SR}: To investigate the superconducting gap symmetry, we performed muon spin rotation and relaxation measurements in the transverse field (TF) configuration. To generate a well-ordered flux line lattice (FLL), the field-cooled method was employed where a magnetic field $H_{c1}$ $<$ $H$ $\ll$ $H_{c2}$ was first applied in the normal state, and then the sample was cooled to low temperature. The asymmetry spectra recorded below and above $T_{c}$ are presented in \figref{TF}(a). The difference in the depolarization rate for the $T$ $<$ $T_{c}$ spectra compared to the above $T_{c}$ can be explained by the presence of an inhomogeneous field distribution due to FLL formation. The weak Gaussian damping for the $T$ $>$ $T_{c}$ spectra corresponds to the relaxation from randomly oriented nuclear dipole fields. The time-domain spectra are well described using oscillatory function multiplied with Gaussian relaxation.
\begin{equation}
 A_{TF}(t) = A~\mathrm{exp} \left (\frac{- \sigma^{2} t^{2}}{2}\right) \mathrm{cos}(\omega t+ \phi),
\label{ATF}
\end{equation}
where $A$ is the asymmetry, $\sigma$ is the relaxation rate of the sample, $\phi$ is the initial phase, and $\omega$ is the precession frequency. The total depolarization of the sample $\sigma$ includes the contribution of FLL in the superconducting state $\sigma_{sc}$ and a small contribution from randomly oriented nuclear dipole moments $\sigma_{nm}$. Figure \ref{TF}(b) shows the fit values of $\sigma$ as a function of the reduced temperature ($T / T_c$) for Re$_{8}$NbTa obtained in the applied field of 100~mT and 400~mT. Due to the formation of FLL in the superconducting state, the field distribution becomes inhomogeneous, resulting in an increase in $\sigma$ below $T_{c}$. The temperature-independent nuclear contribution to the depolarization rates are taken to be $\sigma_{nm}$ = 0.2999(2) and 0.2934(1)~$\mu s^{-1}$ at 100~mT and 400~mT, respectively, equal to the average depolarization rate, $\sigma$, above $T_{c}$.

The total depolarization rate is given by
\begin{equation}
    {\sigma}^2 = {\sigma_{sc}}^2 +{\sigma_{nm}}^2.
\label{sigsc}
\end{equation}
Using \equref{sigsc} and the measured normal state relaxation rate, we extracted the superconducting contribution to the relaxation rate $\sigma_{sc}$. $\sigma_{sc}$ is related to the mean-square inhomogeneity in the field~\cite{fpe},$\langle(\Delta B)^{2}\rangle$ as
\begin{equation}
    {\sigma_{sc}}^{2} = {\gamma_{\mu}}^{2} \langle(\Delta B)^{2}\rangle ,
\end{equation}
where $\gamma_{\mu} = 2\pi \times$135.5~MHz/T is the muon gyromagnetic ratio.

For a triangular Abrikosov vortex lattice, the penetration depth can be calculated from the relaxation rate of FLL using Brandt's formula~\cite{EHB}
\begin{equation}
\label{lambdasc}
\sigma_{sc}(T) = \frac{0.0609 \times \gamma_{\mu} \phi_{0}}{(\lambda^{\mu}(T))^{2}},
\end{equation}
Here, $\phi_{0}$ is the magnetic flux quantum.

The reduced temperature variation of $(\lambda^{\mu})^{-2}$ is shown in \figref{TF}(c). We have used the fitting function for the temperature-dependence of $(\lambda^{\mu})^{-2}$ using London's approximation for a BCS superconductor in the dirty limit~\cite{pene}
\begin{equation}
\frac{\sigma_{sc}(T)}{\sigma_{sc}(0)} = \frac{(\lambda^{\mu}(T))^{-2}}{(\lambda^{\mu}(0))^{-2}} = \frac{\Delta(T)}{\Delta(0)}\mathrm{tanh}\left[\frac{\Delta(T)}{2k_{B}T}\right],
\label{London}
\end{equation}
where $\Delta(T)$ = $\Delta(0)$ tanh[1.82(1.018(($\mathit{T_{c}/T}$)-1))$^{0.51}$], the temperature dependence of the energy gap in the BCS approximation, and $\Delta(0)$ is the energy gap value at 0~K. Our results for this fitting, illustrated in \figref{TF}(c), give $\Delta(0)$ values of 1.20(1)~meV and 1.09(1)~meV at 100~mT and 400~mT, respectively.  We also obtain $\sigma_{sc}(0)$ = 0.4497(9) and 0.3949(5)~$\mu$$s^{-1}$ and the corresponding magnetic penetration depth at 0~K is estimated as $\lambda^{\mu}(0)$ = 4589(7) and 5208(3)~$\text{\AA}$ at 100~mT and 400~mT, respectively.
The excellent fit to \equref{London} is consistent with fully gapped s-wave superconductivity, in agreement with our specific heat results. The normalized superconducting gap value, $\Delta(0)/k_{B}T_{c}$ = 1.88(2) at 100~mT field which further reduced to 1.72(2) at 400~mT, align closely with the value obtained from the specific heat (1.89), suggesting a moderate electron-phonon coupling in the superconducting state of Re$_{8}$NbTa.

\begin{figure} [t!]
\includegraphics[width=0.98\columnwidth,origin=b]{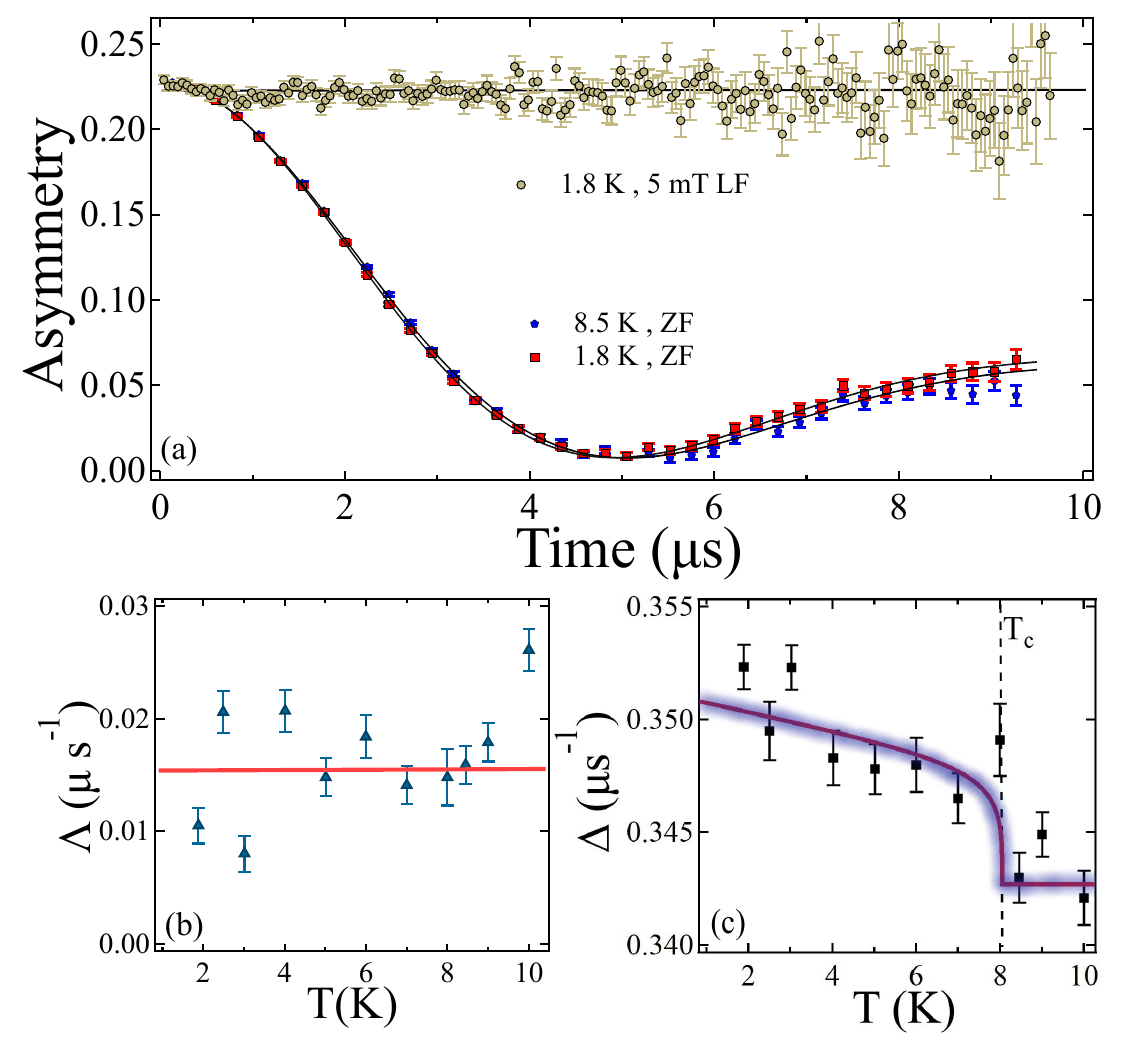}
\caption{\label{ZF}(a) ZF-$\mu$SR spectra collected below (1.8~K) and above (8.5~K) $T_{c}$ and fitted using the Gaussian Kubo-Toyabe function as \equref{AZF}. (b) Temperature dependence of electronic relaxation rate $\Lambda$ with no noticeable change near $T_{c}$. (c) Temperature dependence of nuclear relaxation rate $\Delta$ shows a notable increase below $T_{c}$.}
\end{figure}

\textit{Zero-field $\mu$SR}: Zero-field $\mu$SR measurements were carried out to detect possible time-reversal symmetry breaking (TRSB) in the superconducting state of Re$_{8}$NbTa. Relaxation spectra were collected in the absence of any magnetic field, and two of these spectra recorded below (T = 1.8~K) and above (T = 8.5~K) $T_{c}$ are shown in \figref{ZF}(a). There were no oscillatory components in the spectra, which indicates the absence of any magnetic ordering. In the SC state, a slight enhancement of the relaxation in the superconducting state shown in \figref{ZF}(a) points to the presence of increased internal magnetic fields below $T_{c}$. Asymmetry spectra can be modeled for static and randomly oriented nuclear moments using the Gaussian Kubo-Toaybe (KT) function~\cite{KT}
\begin{equation}
\label{KT}
    G_{KT}(t)= \frac{1}{3}+\frac{2}{3} (1- {\Delta}^{2}t^{2}) e^{\frac{{-\Delta}^{2}t^{2}}{2}} ,
\end{equation}
where $\Delta$ resembles the relaxation rate due to the densely-packed, randomly-oriented nuclear dipoles. To fit our data we used the following function, which is the product of $G_{KT}(t)$ with an exponential function
\begin{equation}
\label{AZF}
    A(t) = A_{i}~G_{KT}(t) e^{-\Lambda t},
\end{equation}
where $A_{i}$ is the initial asymmetry and $\Lambda$ is the exponential relaxation rate. \figref{ZF}(a) shows a small difference that indicates an additional relaxation of the muon spin polarization below $T_{c}$. Our ZF-$\mu$SR data are well fitted using \equref{AZF}. We found that the exponential term remained small and temperature-independent through the superconducting transition as shown in \figref{ZF}(b), while an increase in the Kubo-Toyabe relaxation rate $\Delta$ below $T_{c}$ [see \figref{ZF}(c)] indicates the presence of a small spontaneous magnetic field in the superconducting state. This observation indicates TRSB in Re$_{8}$NbTa and includes it in the family of rhenium-based NCS superconductors with TRSB~\cite{Re6Zr,Re6Hf,Re6Ti}. The distinctive increase in the relaxation rate at $T_{c}$ in Re$_{8}$NbTa, a system characterized by greater disorder than those structurally similar binary alloys~\cite{Re5.5Ta, Re0.82Nb0.18}, indicates a potential role of disorder in TRSB. Similar behavior has been observed in ion-irradiated samples of Ba$_{0.27}$K$_{0.73}$Fe$_{2}$As$_{2}$, which exhibit a clear TRSB signature compared to pristine crystals~\cite{BKFA}. More in-depth investigations are necessary to fully understand the influence of the disorder on TRSB.

We also performed longitudinal field (LF)-$\mu$SR measurements to exclude the possibility of obtaining an increased relaxation from dilute fluctuating impurities. Applying a field of ~5 mT parallel to the muon spin direction at 1.8~K was sufficient to decouple the muon spin polarization from the internal magnetic fields, as evidenced by the essentially flat spectrum in \figref{ZF}(a), indicating that the relaxation seen in ZF primarily corresponds to static internal fields. The magnitude of the internal magnetic field appearing below $T_{c}$ can be estimated~\cite{Re6Ti} from
\begin{equation}
\label{Bint}
B_{int} = \sqrt{2}~\frac{\delta\Delta}{\gamma_{\mu}}
\end{equation}
where $\delta$$\Delta$ is the increase in the Gaussian relaxation rate ($\Delta$) in ZF mode below $T_{c}$, 0.0094(2)~$\mu$$s^{-1}$, which corresponds to the internal field $B_{int}$ = 0.15(1)~G. This value of $B_{int}$ for Re$_{8}$NbTa is comparable to the values found in previously reported NCS rhenium-based TRSB superconductors listed in \tableref{tbl1}. We note that the same order of internal magnetic field corresponding to the change in the exponential relaxation rate was also found in LaNiC$_{2}$~\cite{LNC2}, and La$_{7}$X$_{3}$~\cite{L7I3, L7R3, L7N3}.

\begin{table}[htbp!]
\caption{Internal magnetic field obtained from ZF-$\mu$SR measurement}
\label{tbl1}
\begin{tabular}{c | r} 
Compound & B$_{int}$ (Gauss)\\
\hline\hline
Re$_{8}$NbTa &0.15\\
Re$_{6}$Ti &0.14\\
Re$_{24}$Ti$_{5}$ &0.13\\
Re$_{6}$Zr &0.11\\
Re$_{6}$Hf &0.085\\
\hline
\end{tabular}
\end{table}
\begin{figure}
\includegraphics[width=0.98\columnwidth,origin=b]{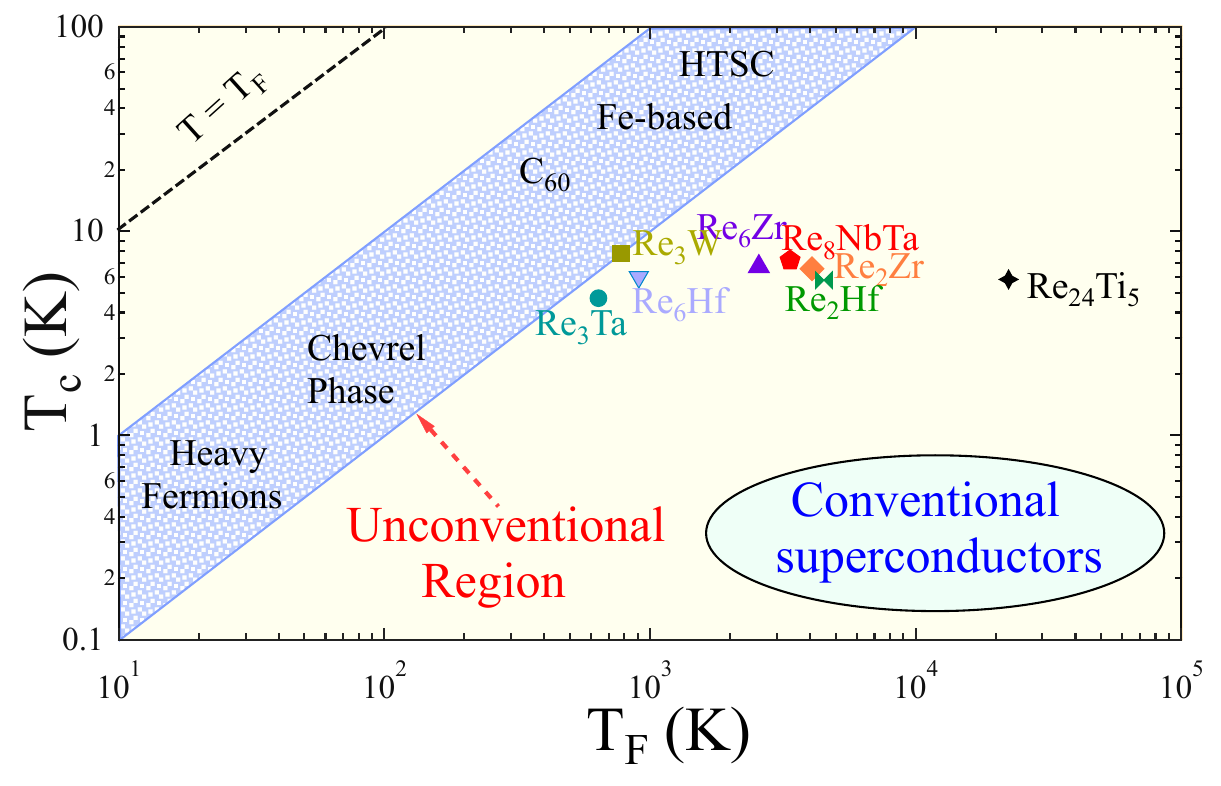}
\caption{\label{umera}Uemura plot between $T_{c}$ and $T_{F}$, Re$_{8}$NbTa is shown as a red pentagon which lies outside the unconventional band.}
\end{figure}
\subsection{Electronic Properties and the Uemura Plot}
To get a deeper understanding of the electronic properties of Re$_{8}$NbTa, we utilized normal and superconducting state parameters derived from both bulk and $\mu$SR measurements. The Hall resistivity ($\rho_{xy}$) at 10~K under the applied magnetic field of $\pm$9~T, measured to estimate the quasiparticle number density (see \figref{rho}(c)). $\rho_{xy}(H)$ is fitted with a linear equation and the slope provides the Hall coefficient $R_{H}$ = $-$ 2.06(1) $\times$ 10$^{-10}$ $\ohm \cdot m T^{-1}$. The negative slope indicates the electron-dominated charge carriers. The quasiparticle number density ($n_{e}$) is related to the Hall coefficient by the expression $R_{H} = -1/e~n_{e}$, which yields, $n_{e}$ = 29.1(2)$\times$10$^{27}$~m$^{-3}$. This enables the estimation of the Fermi wave vector $k_{F}$ = 0.95(1)~$\text{\AA}^{-1}$ using the equation $k_{F}$ = $\left(3\pi^{2}n_{e}\right)^{1/3}$. The effective mass was estimated as $m^{*} = 11.28(7)~m_{e}$ by the relation $m^{*} = \frac{(\hbar k_{F})^{2} \gamma_{n}}{\pi^{2}n_{e} k_{B}}$. Following Drude's model for electrical transport, the mean free path is defined as $l_{e} = v_{F}~\tau$, where $v_{F}=\frac{\hbar~k_{F}}{m^{*}}$, Fermi velocity and average scattering time $\tau = \frac{m^{*}}{n_{e} e^{2}\rho_{0}}$, where residual resistivity $\rho_{0}$ = 108.6(3)~$\mu\Omega$-cm. Using estimated values of $m^{*}$, $n_{e}$, and $\rho_{0}$, we obtain $\tau$ = 0.126(3)$\times$10$^{-13}$~s and $v_{F}$ = 0.98(2)$\times$10$^{5}$~m s$^{-1}$. The mean free path is estimated to be $l_{e}$ = 12.34(1)~\text{\AA}. The BCS coherence length is defined as $\xi_{0}=\frac{0.18 \hbar v_{F}}{k_{B} T_{c}}$ =  186.4(7)~\text{\AA}, higher than the GL coherence length obtained from $H_{c2}(0)$. Since $\xi_{0}\gg l_{e}$, Re$_{8}$NbTa is classified as a superconductor in a dirty limit. The GL penetration depth in dirty limit, at T = 0~K is defined as $\lambda^{dirty}_{GL}(0)=\lambda_{L}(1+\frac{\xi_{0}}{l_{e}})^{1/2}$, where $\lambda_{L}=(\frac{m^{\ast}}{\mu_{0}n_{e}e^{2}})^{1/2}$, the London penetration depth in the dirty limit, which evaluates $\lambda^{dirty}_{GL}(0)$ = 4189(7)~\text{\AA}. Notably, this value aligns closely with $\lambda^{\mu}(0)$ obtained from our TF-$\mu$SR measurement at 100~mT. 
Considering superfluid density $n_{s}$ at T = 0~K is equivalent to $n_{e}$ above $T_{c}$~\cite{LRB2} and assuming a 3D spherical Fermi surface, the Fermi temperature is defined as follows~\cite{LRB2,Tf}
\begin{equation}
k_{B}~T_{F}= \left(\frac{\hbar^{2}}{2 m^{*}}\right) \left(3\pi^{2}n_{s}\right)^{2/3} .
\label{para5}
\end{equation}
\equref{para5} yields $T_{F}$ = 3376(90)~K and 3748(410)~K from bulk and $\mu$SR measurements. The calculated ratio $\frac{T_{c}}{T_{F}}$ = 0.002 places Re$_{8}$NbTa in proximity to other rhenium-based superconductors~\cite{Re6Zr,Re6Hf,Re6Ti,Re3W1,Re3W2,Re3Ta,Re2Hf} within the Umera plot (see Figure \ref{umera})~\cite{Uemura}. All superconducting and normal state parameters of Re$_{8}$NbTa are listed in \tableref{tbl2}.

\begin{table}
\caption{Superconducting and normal state parameters of Re$_{8}$NbTa with Re$_{5.5}$Ta~\cite{Re5.5Ta} and Re$_{0.82}$Nb$_{0.18}$~\cite{Re0.82Nb0.18,ReTRSB}}
\label{tbl2}
\begin{tabular}[b]{l c c c}\hline\hline
Parameters& Re$_{8}$NbTa &Re$_{5.5}$Ta &Re$_{0.82}$Nb$_{0.18}$\\
\hline
$T_{c}$(K)& 7.7(4) &8.0 &8.8\\
$H_{c1}(0)$(mT)& 6.49(4) &3.23 &5.56\\
$H_{c2}(0)$(T)& 17.9(2) &16.47 &17.3\\
$H_{c2}(0)^{P}$(T)& 14.3(2) &14.78 &16.8\\
$H_{c2}(0)^{orb}$(T)& 15.5(1) &9.33 &14.04\\
$\Delta C_{el}/\gamma_{n}T_{C}$& 1.64(2) &2.04 &1.86\\
$\Delta(0)/k_{B}T_{C}$& 1.89(1) &1.99 & 1.83\\
$\gamma_{n}$(mJ mol$^{-1}$K$^{-2}$)& 44.5(4) &25.3 &53.5\\
$\theta_{D}$(K)& 303(1) &310 &383.1\\
$\lambda_{e-ph}$& 0.73(2) &0.73 &0.73\\
$\xi_{GL}(0)$(\text{\AA})& 42.8(7) &45 &43.6\\
$\lambda_{GL}(0)$(\text{\AA})& 3371(2) &4949 &3625\\
$k_{GL}$& 78.6(2) &111 &83.14\\
$\lambda^{\mu}(0)$(\text{\AA})& 4589(7) &4226 &3573\\
$\lambda^{dirty}_{GL}(0)$(\text{\AA})& 4189(5) &- &-\\
$v_{F}$($10^{5}$ m$s^{-1}$)&0.98(2) &12.17 &-\\
$n$($10^{27}$ $m^{-3}$)&29.1(2) &2.8 &-\\
$T^{bulk}_{F}$(K)&3376(90) &2040 &-\\
$T^{\mu}_{F}$(K)&3748(410) &- &-\\
${T_{c}}/{T_{F}}$&0.002(1) &0.0038 &-\\
${m^{*}}/{m_{e}}$&11.28(7) &4.15 &-\\
\hline\hline
\end{tabular}
\par\medskip\footnotesize
\end{table}

\section{Conclusion}
In conclusion, we have synthesized a new ternary noncentrosymmetric bulk superconductor Re$_{8}$NbTa having a $T_{c}$ of 7.7(4)~K with an upper critical field surpassing the Pauli limiting field. The superconducting gap $\Delta(0)/k_{B}T_{c}$ = 1.89, and the electron-phonon coupling strength $\lambda_{e-ph}$ = 0.73, point to a robust electron-phonon interaction suggesting moderately coupled superconductivity. Temperature-dependent electronic specific heat and TF-$\mu$SR measurements confirm the existence of a nodeless superconducting gap. Additionally, ZF-$\mu$SR results unveil the breaking of the time-reversal symmetry in the superconducting state, characterized by an internal magnetic field of 0.15~G. Our result suggests the influence of disorder on the emergence of time-reversal symmetry breaking in the superconducting state of Re$_{8}$NbTa, while the role of the critical Re-percentage appears to be insignificant. Nevertheless, further comprehensive investigations on new Re-based superconductors are necessary to establish a conclusive link between disorder, crystal structure, anti-symmetric spin-orbit coupling (ASOC), and time-reversal symmetry breaking (TRSB) in Re-based superconducting alloys.

\section{Acknowledgment}
R.K.K. acknowledges the UGC Government of India for the SRF fellowship. R.P.S. thanks SERB, Government of India, for the Core Research Grant CRG/2019/001028. Work at McMaster University received support from NSERC, Canada.

\end{document}